\documentclass[onecolumn, twoside, 11pt]{article}

\usepackage{parskip}
\usepackage{amsmath,amsfonts,amssymb}
\usepackage{graphicx}
\usepackage{xcolor}
\usepackage{arydshln}

\numberwithin{figure}{section}

\usepackage{tikz}
\tikzset{  highlight/.style={rectangle,rounded corners,fill=red!15,draw,fill opacity=0.5,thick,inner sep=0pt}
}

\usepackage{pgfplots}
\pgfplotsset{compat=newest}
\pgfplotsset{plot coordinates/math parser=false}
\newlength\figureheight
\newlength\figurewidth

\usetikzlibrary{arrows,intersections}

\usetikzlibrary{calc,decorations.pathmorphing,patterns}

\makeatletter
\pgfdeclaredecoration{penciline}{initial}{
    \state{initial}[width=+\pgfdecoratedinputsegmentremainingdistance,auto corner on length=0.25mm,]{
        \pgfpathcurveto        {            \pgfqpoint{\pgfdecoratedinputsegmentremainingdistance}
                            {\pgfdecorationsegmentamplitude}
        }
        {        \pgfmathrand
        \pgfpointadd{\pgfqpoint{\pgfdecoratedinputsegmentremainingdistance}{0pt}}
                        {\pgfqpoint{-\pgfdecorationsegmentaspect\pgfdecoratedinputsegmentremainingdistance}                                        {\pgfmathresult\pgfdecorationsegmentamplitude}
                        }
        }
        {        \pgfpointadd{\pgfpointdecoratedinputsegmentlast}{\pgfpoint{0.5pt}{0.5pt}}
        }
    }
    \state{final}{}
}
\makeatother

\usetikzlibrary{shapes,arrows}

\tikzstyle{block} = [draw, fill=white, rectangle, minimum height=3em, minimum width=3em]
\tikzstyle{largeblock} = [draw, fill=white, rectangle, minimum height=3em, minimum width=3em]
\tikzstyle{smallblock} = [draw, fill=white, rectangle, minimum height=3em, minimum width=3em]
\tikzstyle{input} = [coordinate]
\tikzstyle{sum} = [draw, fill=white, circle, minimum size=2pt, inner sep=2pt, label={center:\tiny$+$}]
\tikzstyle{node} = [draw, fill=black, circle, minimum size=2pt, inner sep=0pt]
\tikzstyle{output} = [coordinate]
\tikzstyle{pinstyle} = [pin edge={to-,thin,black}]
\tikzstyle{line} = [draw, -latex']

\usepackage{amsthm}
\usepackage{thmtools}

\declaretheoremstyle[  within=section,            notefont=\normalfont\itshape,  ]{definitionstyle}

\declaretheorem[style=definitionstyle,name=Exercise]{exercise}
\declaretheorem[style=definitionstyle,name=Lemma]{lemma}
\declaretheorem[style=definitionstyle,name=Theorem]{theorem}

\newcommand{\norm}[1]{\lVert#1\rVert}
\newcommand{\vbar}{\Big\vert}
\newcommand{\inprod}[2]{\left<{#1},{#2}\right>}
\newcommand{\mean}[1]{\overline{#1}}

\newcommand\vf{\mathbf{f}}
\newcommand\vg{\mathbf{g}}

\newcommand\vu{\mathbf{u}}
\newcommand\vv{\mathbf{v}}

\newcommand\vx{\mathbf{x}}

\newcommand\vz{\mathbf{z}}

\newcommand\for{\mathbf{\phi}}
\newcommand\res{\mathbf{\psi}}

\newcommand\vhf{\hat{\mathbf{f}}}

\newcommand\vhu{\hat{\mathbf{u}}}

\newcommand\vhz{\hat{\mathbf{z}}}

\newcommand\hp{\hat{p}}
\newcommand\tp{\tilde{p}}

\newcommand\vtu{\tilde{\mathbf{u}}}

\newcommand\vtz{\tilde{\mathbf{z}}}

\newcommand\mL{\mathbf{L}}
\newcommand\mR{\mathbf{R}}
\newcommand\mU{\mathbf{U}}
\newcommand\mV{\mathbf{V}}
\newcommand\mSigma{\Sigma}

\newcommand\N{\mathbb{N}}
\newcommand\R{\mathbb{R}}

\usepackage{listings}
\title{Elements of resolvent methods in fluid mechanics: notes for an introductory short course v0.3}
\author{A S Sharma\\ a.sharma@soton.ac.uk\\ University of Southampton}

\begin{document}
\maketitle

\section{Introduction}

This is a collection of notes for part of a short course on modal methods in fluid mechanics held at DAMTP, University of Cambridge, in the summer of 2019.
These notes in particular are meant to introduce the reader to resolvent analysis as it is currently used in fluid mechanics. Most of the papers on the topic assume a level of knowledge a bit beyond that of the average beginning PhD student, so there is a need for some introductory material to get new students up to speed quickly. These notes are a step towards providing such material and will serve as a base from which to explore the literature on the topic.
The presentation assumes a good working knowledge of Fourier transforms and linear algebra, some familiarity with the incompressible Navier-Stokes equations, and not much else. Some experience with state space systems from an introductory course in control is beneficial. In most cases, rigour and technical detail have been elided in order not to obscure the central point. Inevitably, there will be mistakes in the notes and I would be grateful to be informed of these by email.

The method of analysis described in what follows arose from a desire to have a systematic and well-founded way to form `quick and dirty' approximations to turbulent Navier-Stokes flows from the equations themselves (that is, as far as possible without recourse to simulation or experimental data). It was hoped that such approximations would successively approach the original equations as the detail of the approximation was increased. Fast and simple calculations would then enable the kind of parametric control studies that are expensive with direct numerical simulation.

This kind of approach was inspired by the successful model reduction methods of modern linear control theory, such as balanced truncation. Unfortunately, the existing methods of the time were designed for linear systems, or nonlinear systems that could sensibly be linearised around an operating point. Although many researchers had long practised looking at linear operators formed around the mean flow, it was not then clear to me what it was that was actually being calculated; the classical linearisation theorem taught to undergraduates explains the correspondence between a nonlinear system and its locally valid linearisation around an equilibrium. In contrast, turbulent flows are far from equilibrium, the turbulent mean is not an equilibrium point in phase space, and the turbulent fluctuations are large.

This dissatisfaction ultimately resulted in the present analysis. If it makes sense to speak of lineage in this context, one may draw a line back through the pseudospectra insights of Trefethen and coworkers \cite{Trefethen.ea:1993}, and the laminar resolvent based work arising from the control theory community \cite{Jovanovic.Bamieh:2005}. Inevitably, this view and the presentation that follows is my own individual perspective.

These notes begin with an introduction to the singular value decomposition and its operator counterpart, the Schmidt decomposition.
A general formulation of the resolvent decomposition is then introduced. A brief discussion of the interpretation as a nonlinear feedback loop is given. The methodology is then applied to the Navier-Stokes equations.

\section{The singular value decomposition}

The singular value decomposition (SVD) is a particular matrix factorisation that has very useful properties. It is widely used in data and model reduction because it solves the problem of finding the optimal approximation of a linear operator.
Since we will be using it extensively, we now review some of its most important properties.
In this section, vectors will be represented by lowercase letters, matrices by uppercase, and the conjugate transpose of $A$ by $A^*$.

\begin{figure}
    \[
        \underbrace{
            \left[
                \begin{array}{*4{c}}
                    \cdot & \cdot & \cdot & \cdot \\
                    \cdot & \cdot & \cdot & \cdot \\
                    \cdot & \cdot & \cdot & \cdot 
                \end{array}
        \right]
    }_{M}
    = \underbrace{
        \left[
            \begin{array}{*3{c}}
                \cdot & \textcolor{magenta}{\cdot} & \textcolor{cyan}{\cdot} \\
                \cdot & \textcolor{magenta}{\cdot} & \textcolor{cyan}{\cdot} \\
                \cdot & \textcolor{magenta}{\cdot} & \textcolor{cyan}{\cdot}  
            \end{array}
            \right]
        }_{U}
    \underbrace{
        \left[
            \begin{array}{ccc:c}
                \sigma_1 &   &   &     \\
                & \textcolor{magenta}{\sigma_2}   &   &   \\
                &   & \textcolor{cyan}{\sigma_3}   &   
            \end{array}
    \right]
    }_{\Sigma}
    \underbrace{
        \left[
            \begin{array}{*4{c}}
                \cdot & \cdot & \cdot & \cdot \\
                \textcolor{magenta}{\cdot} &\textcolor{magenta}{\cdot} &\textcolor{magenta}{\cdot} &\textcolor{magenta}{\cdot} \\
                \textcolor{cyan}{\cdot}  & \textcolor{cyan}{\cdot}  & \textcolor{cyan}{\cdot}  & \textcolor{cyan}{\cdot} \\ 
                \hdashline
                & & &
            \end{array}
        \right]
    }_{V^*}
    \]
    \caption{The structure of the singular value decomposition with $n>m$. The linears in $\Sigma$ and $V^*$ represent the reduced SVD (see Section \ref{sec:truncated svd})}
    \label{fig:svd}
\end{figure}

\begin{lemma}
    Let $M$ be a complex $m \times n$ matrix. 
    The decomposition
    \begin{equation}
        M = U \Sigma V^*
        \label{eq:svd}
    \end{equation}
    always exists, where $U$ is an $m \times m$ complex matrix, $V$ is an $n \times n$ complex matrix, $\Sigma$ is a $m \times n$ real and diagonal matrix with elements $\Sigma_{ii} = \sigma_i$ and
    \(\sigma_1 \geq \sigma_2 \geq \ldots.\)
    The $\sigma_i$ are called the singular values and \eqref{eq:svd} is called the singular value decomposition of $M$.
Matrices $U$ and $V$ are unitary, $U U^* = U^* U = I_m$ and $V V^* = V^* V = I_n$.
\end{lemma}

From the singular value decomposition, we can make the following observations.
Since $U$ and $V$ are unitary, the rank of $M$ is equal to the number of nonzero singular values. Notice that the inverse of a unitary matrix is its conjugate transpose.
The decomposition is unique up to a constant complex multiplicative factor on each basis and up to the ordering of the singular values. 
That is, if $U\Sigma V^*$ is a singular value decomposition, so is $(e^{i\theta} U) \Sigma (V^* e^{-i\theta})$.
The columns of $V$ and $U$ that span the space corresponding to any exactly repeating singular values may be combined arbitrarily.
The structure of the matrix decomposition is illustrated in figure \ref{fig:svd}.

\subsection{The maximum gain problem and its relationship with norms}
\label{sec:maximum gain}
\begin{figure}[h]
    \centering
    \includegraphics[width=0.75\textwidth]{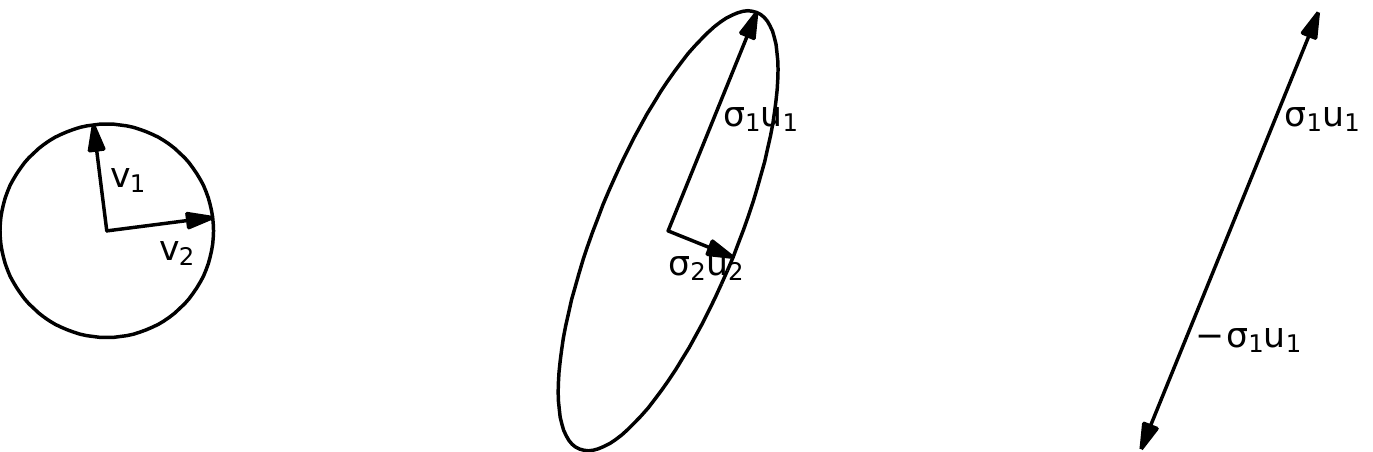}
    \caption{Mapping of the unit circle ($\norm{a}=1$, left) to an ellipse ($Ma$, centre) and mapping of $Mv_1$ to $\sigma_1 u_1$ (right). If we imagine the locus of points of $a$ with unit length being drawn on a rubber sheet, the effect on $M$ is to rotate and stretch the sheet. The amount of stretching in each direction is given by each singular value, and the directions by the singular vectors. }
    \label{fig:svd circle}
\end{figure}

It is helpful to think of $M$ as an operator mapping a complex vector in the domain of $M$ to another in the range of $M$.
The columns of $V$, $v_i$, provide a basis which spans the domain.
The singular value decomposition of $M$ can be written in terms of the vectors of $U$ and $V$,
\begin{equation}
    M = U \Sigma V^* = \sum_{i=1}^m \sigma_i u_i v_i^*.
\end{equation}
Since $V$ is unitary, $v_i^* v_j = \delta_{ij}$, so applying $M$ to $v_j$ gives
\begin{equation}
    M v_j = \sum_{i=1}^m \sigma_i u_i v_i^* v_j = \sigma_j u_j.
\end{equation}
Since $V$ provides a basis for the domain of $M$, any vector $a$ in the domain of $M$ can itself be expressed in terms of a weighted sum of columns of $V$.
That is, expressing $a$ as
\[a = \sum_{i=1}^{n} v_i c_i\]
gives
\begin{eqnarray*}
    Ma &= \sum_{i=1}^m u_i \sigma_i v_i^* a \\
       &= \sum_{i=1}^m u_i \sigma_i c_i.
\end{eqnarray*}
We may then pose the question, what is the maximum amplitude of `output' for a given `input' amplitude? This is achieved with the input parallel to $v_1$, with a gain of $\sigma_1$.
So,
\[
    \sigma_1 = \max_{a \neq 0} \frac{\norm{M a}}{\norm{a}}
\]
is achieved with $a / \norm{a} = v_1$. Any other choice of $a$ that is not parallel to $v_1$ would achieve an inferior gain.
This is illustrated in figure \ref{fig:svd circle} for $a$ of unit length. $M$ maps a circle (ball) of unit radius to an ellipse (hyperellipse). The singular values are the major and minor axes of the ellipse.

\subsection{The low-rank approximation of matrices}
\label{sec:truncated svd}
For a non-square or rank-deficient square matrix, some of the singular values will be zero. In this case, the reduced SVD can be defined where the columns of $U$ or $V$ relating to the zero singular values, and the corresponding entries of $\Sigma$, can be truncated with the decomposition remaining exact. In this case, though, $U$ (or $V$) will not be unitary because the columns associated with the null space of $M$ will have been truncated. This is illustrated in Figure \ref{fig:svd}, where the truncated columns of $\Sigma$ and $V$ are separated from the rest of the matrix by dotted lines.

Since these matrices often arise from numerical calculations, it is natural to ask what to do with singular values that are approximately zero within some defined threshold. If these are truncated, the decomposition forms an approximation to $M$.

Where $M$ is approximated by its SVD expansion truncated to order $r$,
\begin{equation}
    M \simeq M_r = \sum_{i=1}^r u_i \sigma_i v_i^*,
\end{equation}
it can be seen that the approximation error is equal to the rest of the expansion (the `tail'),
\begin{align}
    Ma - M_r a &= \sum_{i=r+1}^{m} u_i \sigma_i v_i^* a,
\end{align}
and so is bounded,
\begin{align}
    \norm{Ma - M_r a} & \leq \sigma_{r+1} \norm{a}.
\end{align}

\begin{figure}
    \centering
    \includegraphics[width=0.8\columnwidth]{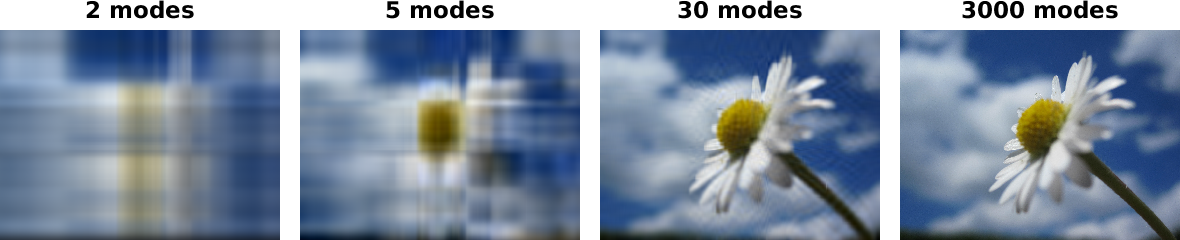}
    \vspace{1em}
    \includegraphics[width=0.8\columnwidth]{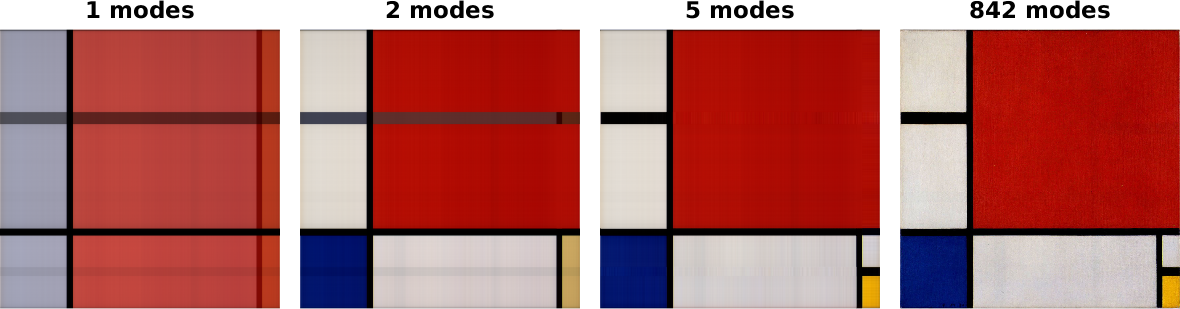}
    \caption{The singular value decomposition applied to image reconstruction, showing progressively higher rank approximations of an image. The upper image series requires a high number of modes to capture the detail. The lower image series shows that the Mondrian picture is well-approximated by a very low-rank projection.}
    \label{fig:svd images}
\end{figure}

The effect of rank reduction can be seen by looking at the following Matlab code extract, which applies the SVD to images.
\begin{lstlisting}
[U, S, V] = svd(img);
aprox_img = U(:,1:r) * S(1:r,1:r) * V(:,1:r)';
\end{lstlisting}
The output for a pair of sample images is shown in Figure \ref{fig:svd images}

\subsection{The pseudo-inverse}
For reference, we briefly mention the pseudo-inverse here. A better and more detailed introduction (with proofs) is presented in \cite{Strang}.
The inverse of a matrix $M$ in terms of its SVD is simply
\begin{eqnarray*}
    M      =& U \Sigma V^* \\
    M^{-1} =& V \Sigma^{-1} U^*.
\end{eqnarray*}
Clearly $\Sigma^{-1}$ only exists if $M$ is full rank and square. Otherwise, we define the pseudo-inverse (or Moore-Penrose inverse) via the reduced SVD,
\begin{eqnarray*}
    M^{+} &= V_r \Sigma_r^{-1} U_r^*
\end{eqnarray*}
where $\Sigma_r$ is the truncation of $\Sigma$ to remove all zero singular values, so that $\Sigma_r$ is invertible.
$M^+$ is a one-sided inverse (which side depends on whether $m > n$; for example, if $m > n$, then $M M^+ = I_n$).

Consider the under- or over-determined linear system of equations
\begin{equation}
    Mx = b.
    \label{eq:least squares}
\end{equation}
A least squares solution $x'$ that minimises $\norm{Mx - b}$ is given by
\[x' = M^+ b.\]
In the case where $M$ does not have linearly independent columns, the solution $x'$ (of all possible solutions) given by the pseudoinverse is the solution that has the minimum length.

\subsection{The singular value decomposition for linear operators}
The matrix SVD has a direct analogy for linear operators on Hilbert spaces, called either the Schmidt decomposition or sometimes also the singular value decomposition.
The reader not used to dealing with functions in and operators on Hilbert spaces may rest assured that the situation is conceptually very similar to the matrix case.
A good and detailed reference is \cite{Young:1988}. The following statement is equivalent to that in \cite{Curtain.Zwart:1995}.
In the following, $\inprod{\cdot}{\cdot}_X$ represents the inner product on the space $X$.

\begin{theorem}
    If $T : X \rightarrow Y$ is a compact (bounded, linear) operator, where $X$ and $Y$ are Hilbert spaces, then $T$ has the following representation:

    \begin{equation}
        T x = \sum_{i=1}^{\infty} \sigma_i \inprod{x}{\phi_i}_X \psi_i,
        \label{eq:Schmidt decomposition}
    \end{equation}
for some $x \in X$ where the set $\{\phi_i\}$ and the set $\{\psi_i\}$ are the eigenvectors of $T^*T$ and $TT^*$ respectively, and $\sigma_i \geq 0$ are the square roots of the eigenvalues. The $\{\phi_i\}$ and $\{\psi_i\}$ form an orthonormal basis for $X$ and $Y$ respectively (so $\inprod{\psi_i}{\psi_j}_Y = \delta_{ij}$ and so on). A pair $(\phi_i, \psi_i)$ is a Schmidt pair of $T$ with an associated singular value $\sigma_i$ and \eqref{eq:Schmidt decomposition} is the Schmidt decomposition of $T$.
Further, $T$ is bounded with norm $\sigma_1$, so $\norm{Tx} \leq \sigma_1 \norm{x}$.
\end{theorem}
Fortunately, many useful properties and much of the intuition arising from the simpler matrix SVD carry over to the operator case.
The most obvious difference is that since $X$ and $Y$ are function spaces, there can be infinitely many singular values.
Even then, $T$ can still be approximated by a finite-rank operator with bounded error, in a manner analogous to the matrix case. There are a few points to be aware of in a numerical implementation. Since $T$ is a mapping between Hilbert spaces, inner products will have been defined for $X$ and $Y$. Therefore, care must be taken to be sure that any discretisation preserves the appropriate inner product in the form of whatever mass matrix, grid weighting or similar is appropriate. This is often a tedious step and not explicitly outlined in papers.

\subsection{Further reading}
The singular value decomposition is an extremely well established and widely used piece of mathematics. For the matrix case, the reader may find the presentations in \cite{Strang} or \cite{Trefethen.Bau:1997} useful and insightful.
The practical application in a control setting is well explained in \cite{Green.Limebeer:1995}.
The operator case is thoroughly covered in \cite{Young:1988} and, in an infinite-dimensional linear systems setting, introduced in \cite{Curtain.Zwart:1995}.

\subsection{Exercises}
\begin{exercise}
    Determine the relationship between the SVD of $M$ and the eigenvalues and eigenvectors of $M M^*$ and of $M^* M$.
\end{exercise}
\begin{exercise}
    Find the singular values of $AM$ where $A$ is a unitary matrix.
\end{exercise}
\begin{exercise}
    Find the minimum singular value of $\lambda I - M$ where $\lambda$ is an eigenvalue of $M$.
\end{exercise}
\begin{exercise}
    For the operator case, and using the eigenvalues and eigenvectors of $T^*T$, show that $\norm{Tx} \leq \sigma_1 \norm{x}$.
\end{exercise}

\section{Resolvent analysis of dynamical systems}

Resolvent analysis in various forms has emerged as a useful tool in fluid dynamics. Here, we present the type introduced in \cite{McKeon.Sharma:2010} which is applicable to a broad range of systems, including nonlinear systems far from a steady equilibrium, such as turbulence. To prevent the general idea from being obscured by details involving fluid dynamics, we start with the general case, then specialise to the Navier-Stokes equations. The basic idea presented here is relatively simple and can be understood using just Fourier transforms, linear algebra and some familiarity with state space.

The approach differs from classical linearisation in the following way. A linearisation would typically proceed by doing the Taylor expansion about an equilibrium and assuming the perturbations around it to be small, leading us to neglect higher-order terms in the perturbations. This allows some qualitative statements to be made about the region near the equilibrium. Clearly, it is convenient to expand around a point where the equation associated with that point has no time derivatives; this is true at an equilibrium but is also true at the mean. In contrast, although in our case $\frac{d}{dt}{\mean{\vz}}$ is zero, the nonlinear terms are not small, so are kept.

We begin by illustrating the idea using a dynamical system with a finite-dimensional state space.
Let the state at time $t$ be $\vz(t)$, let $\vz(t) \in \R^n$, and the dynamics be given by a function $\vg$, $\vg: \R^n \rightarrow \R^n$;
\begin{align}
    \frac{d}{dt}{\vz}(t) &= \vg\left( \vz(t) \right).
    \label{dynamics}
\end{align}

The aim is to get some sense of what the dynamics of the system look like without having to time-integrate it. The approach is to put the system into a form where we can apply tools from linear algebra, even though the system is nonlinear. To do this, we will go into the frequency domain. Once there, a linear operator arising from an expansion around the mean is formed. The most amplified directions of this operator are found, and are assumed to be excited by the remaining nonlinear terms in the expansion. These most amplified directions represent most favoured motions in the state space for any given frequency.

In the following, we assume we know the long-time average of the state, $\mean{\vz}$.\footnote{This is obviously a weakness of the method if it needs to be found by time integration, but in some cases it may be known by other means such as experiment.}
The Taylor expansion of \eqref{dynamics} about $\mean{\vz}$ is
\begin{align}
    \frac{d}{dt}{\vz}(t) &= \frac{d}{dt}\mean{\vz} + \frac{d}{dt}({\vz}(t) - \mean{\vz}) \\
                       &= \vg(\mean{\vz}) + \frac{\partial \vg}{\partial \vz} \vbar_{\bar{\vz}}  (\vz(t) - \mean{\vz}) + \mathrm{H.O.T.} \\
    &= \mL (\vz(t) - \mean{\vz}) + \vf(t)
    \label{eq:expansion}
\end{align}
where $\vf$ collects together the terms nonlinear in the fluctuations $(\vz(t) - \mean{\vz})$ and the constant term $\vg(\mean{\vz})$.
$\mL$ is the Jacobian of $\vg$ about $\bar{\vz}$. This expansion is essentially just a change of variables with the new origin being the mean. This splitting into mean and fluctuations is basically a Reynolds decomposition.
The expansion around $\mean{\vz}$ is then Fourier transformed. This involves the Fourier transform of $\vz$,\footnote{Really, this integral diverges, because our signal is not bounded.}
\begin{align}
    \vhz(\omega) &= \int_{-\infty}^{\infty} e^{-i \omega t} \vz(t) \ dt
\end{align}
with inverse transform
\begin{align}
    \vz(t) &= \frac{1}{2 \pi }\int_{-\infty}^{\infty} e^{i \omega t} \vhz(\omega) \ d\omega,
    \label{eq:ift}
\end{align}
and of $\vf$,
\begin{align}
    \vhf(\omega) &= \int_{-\infty}^{\infty} e^{-i \omega t} \vf(t) \ dt.
\end{align}
The mean, $\mean{\vz}$, is closely related to the $\omega=0$ component. It can be verified by substitution into \eqref{eq:ift} that
\begin{align}
    \vhz(\omega=0) &= 2 \pi \mean{\vz} \delta(\omega),
    \label{eqn:mean}
\end{align}
with $\delta$ being the Dirac delta, and the equation corresponding to $\omega=0$ is therefore the mean equation.

To proceed, we integrate the expansion \eqref{eq:expansion} against a chosen frequency $\omega$.
At any particular $\omega \neq 0 $ we then have
\begin{equation}
    i \omega  \vhz(\omega) = \mL \vhz(\omega) + \vhf(\omega).
\end{equation}
which can be rearranged as
\begin{equation}
    \vhz(\omega) = \left( i \omega I - \mL \right)^{-1} \vhf(\omega).
\end{equation}
The operator $\mR(\omega) := \left( i \omega I - \mL \right)^{-1}$ which maps the Fourier coefficient of the nonlinear excitation, $\vhf$ to the Fourier coefficient of the state, $\vhz$, is the resolvent operator (matrix, in this case) of $\mL$. It is essentially a transfer function from $\vhf$ to $\vhz$.

It is worth pausing here to think about what has been done.
The benefit of this approach is that it brings us to a point where the familiar and powerful tools of linear algebra can be applied, despite the actual dynamics being highly nonlinear.
The nonlinear terms in $\vhf$, rather than being discarded, are essential and act to excite the state. Since the effect of these terms is filtered by the linear part of the dynamics, we will be able to say something about how the filtering affects the state.
The tool of choice in this situation is the singular value decomposition, which we will apply to the resolvent.

\subsection{The singular value decomposition of the resolvent operator}
The question at hand is what typical motions in state space are to be expected. From the preceding discussion we have shown that a matrix $\mR$ maps some unknown vector $\vhf$ to the state's Fourier coefficient $\vhz$. This matrix can be found from the governing equations and is the resolvent of the Jacobian formed around the long-time average state. We shall seek to find a basis in which to express $\vhz$ as accurately as possible and with as few coefficients as possible, but without knowing $\vhz$. This means we have to guess what $\vhz$ is going to look like. To do this, we find the optimal approximation to $\mR(\omega)$. Since $\mR(\omega)$ is just a matrix, this is given by the singular value decomposition at each $\omega$,
\begin{equation}
    \mR(\omega) = \mU(\omega) \mSigma(\omega) \mV^*(\omega).
\end{equation}
This decomposition at any particular frequency $\omega$ induces a basis in which to express $\vhz$, which is the left singular vectors (the columns of $\mU(\omega)$).
In the absence of any knowledge about $\vhf$, we may truncate this basis to order $r$ with the hope that the relative smallness of the trailing singular values of $\mR(\omega)$, that is, $\sigma_{j}(\omega)$ with $j>r+1$, act to preclude any component of $\vhz$ in the directions $\vhu_{j}$, with $j>r+1$.

The Fourier coefficients of the state can be expanded in terms of the left singular vectors of $\mR(\omega)$,
\begin{align}
    \vhz(\omega) &= \mR(\omega) \vhf(\omega) \\
                 &= \sum_{i=1}^r \vu_i(\omega) \sigma_i(\omega) \vv_i^*(\omega) \vhf(\omega) \\
                 &= \sum_{i=1}^r \vu_i(\omega) \sigma_i(\omega) c_i(\omega).
\end{align}
The truncation to order $r$ defines an optimally reduced space that $\vhz(\omega)$ inhabits. The $c_i$ are the coefficients obtained by the projection of $\vhf(\omega)$ onto the right singular vectors, i.e.~$c_i = \vv_i^*(\omega) \vhf(\omega)$. To fix the coefficients $c_i$ and calculate a specific trajectory $\vz$ would require invoking the nonlinear term.

\subsection{Exercises}
In \cite{Trefethen.ea:1993}, the resolvent operator formed around the laminar flow is suggested as a model for transition. Many of the ideas presented in these notes are introduced. In the same paper, the simple conceptual model system is proposed,
\begin{equation}
    \frac{d\vu}{dt}(t) = 
    \left[
    \begin{array}{cc}
        -1/\rho & 1 \\
        0 & -2/\rho
    \end{array}
    \right]
    + \norm{\vu}
    \left[
    \begin{array}{cc}
        0 & -1\\
        1 & 0
    \end{array}
    \right]
    \vu.
\end{equation}

\begin{exercise}
    Find the eigenvalues of the linear operator in the model system.
\end{exercise}

\begin{exercise}
    Using the same linear operator, form the resolvent operator for $\rho=25$. 
    Plot the leading singular value as it varies with $\omega$.
    What happens to the leading singular value when there is an eigenvalue close to the imaginary axis?
\end{exercise}

\section{Nonlinear feedback, solutions to the nonlinear system and the Lur'e decomposition}
\label{sec:Lur'e}

In this section, we seek to understand the global behaviour when both nonlinear and linear parts of the system co-exist. Some authors eschew this interpretation and instead choose to model $\vhf$ statistically using some prior knowledge obtained through other means. Both approaches are reasonable.

\begin{figure}
    \centering
    \begin{tikzpicture}[auto, node distance=2cm,>=latex']
        \node [block] (H) {$H$};
        \node [output, right of=H] (out) {};
        \node [input, left of=H] (in) {};
        \draw [->] (in) -- node[above] {$\vf$} (H);
        \draw [->] (H) -- node[above] {$\vtz$} (out);
    \end{tikzpicture}
    \caption{A block diagram representation of the open-loop relation $H$.}
    \label{fig:H}
\end{figure}
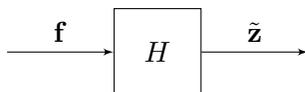

Consider the dynamics in \eqref{eq:expansion},
\begin{equation}
    \frac{d}{dt} \vtz(t) = \mL \vtz(t) + \vf(t)
    \label{eq:linear dynamics}
\end{equation}
where $\vtz$ is defined to be the fluctuations at time $t$, $\vtz:=\vz(t) - \mean{\vz}$.
This system can be interpreted as a linear system with external forcing $\vf(t)$. As such, it can be integrated from some initial condition to solve for $\vtz(t)$.
We have dealt with $\vtz(t)$ (the state at some time $t$) as a vector in $\R^n$, but we can also talk of the solution being a function $\vtz:\R \rightarrow \R^n$ which maps a real number ($t$) to a point in $\R^n$ (the state at $t$).\footnote{Care should be taken to understand that this function represents a whole trajectory. A confusion often arises at this point because of the common practice of omitting the argument $t$ when discussing the instantaneous state, thus leading to a confusion between $\vtz$ and $\vtz(t)$.} This trajectory can be related to the equivalent forcing $\vf$ by a linear operator $H$,
\begin{equation}
    \vtz = H \vf.
\end{equation}
That is to say, $H$ maps a whole history of $\vf(t)$ (the signal $\vf$ in the space of such functions) to a whole history of $\vtz(t)$ (a trajectory in the space of such functions).
As such, if $\vf$ was known, applying $H$ would result in $\vtz$, which could give the state at any time $t$ by evaluating $\vtz(t)$. $H$ could be calculated the time-integration of \ref{eq:linear dynamics} with an initial condition.
This is depicted in Figure \ref{fig:H}.

The counterpart is the nonlinear relation
\begin{equation}
    \vf = N(\vtz)
\end{equation}
where $N$ calculates the function $\vf$ from the state trajectory $\vtz$. Again, the nonlinear term $\vf(t)$ is calculated by supplying $t$ as an argument to $\vf$.
In the special case that $N$ is memoryless, $\vf(t)$ can be calculated instantaneously from $\vtz(t)$ without knowing the whole time history.

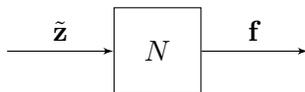
\begin{figure}
    \centering
    \begin{tikzpicture}[auto, node distance=2cm,>=latex']
        \node [block] (N) {$N$};
        \node [output, right of=H] (out) {};
        \node [input, left of=H] (in) {};
        \draw [->] (in) -- node[above] {$\vtz$} (N);
        \draw [->] (N) -- node[above] {$\vf$} (out);
    \end{tikzpicture}
    \caption{A block diagram representation of the open-loop nonlinear relation $N$.}
    \label{fig:N}
\end{figure}

Similarly to the open loop relation $H$, $N$ maps a signal $\vtz$ to an $\vf$.
As such, if $\vtz$ was known, the whole history of $\vf(t)$ could be reproduced finding $\vf$ and then evaluating at $t$.
This is depicted in Figure \ref{fig:N}.

\begin{figure}
    \centering
    \begin{tikzpicture}[auto, node distance=2cm,>=latex']
        \node [block] (N) {$N$};
        \node [block, below of=N] (H) {$H$};
        \node [output, right of=N] (Nout) {};
        \node [input, left of=N] (Nin) {};
        \draw [->] (Nin) -- node[above] {$\vtz$} (N);
        \draw [-] (N) -- node[above] {$\vf$} (Nout);
        \draw [-] (H) -| node[above] {} (Nin);
        \draw [->] (Nout) |- node[above] {} (H);
    \end{tikzpicture}
    \caption{A block diagram representation of the closed-loop system consisting of $H$ and $N$.}
    \label{fig:CL}
\end{figure}
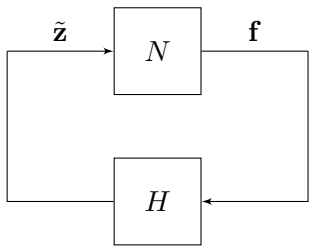
A solution to the original equations is a pair $(\vf,\vtz)$ that simultaneously satisfies both open-loop relations.
The `closed-loop' is such an arrangement, depicted in Figure \ref{fig:CL}. It should be noted that since $H$ is linear, fixing the amplitudes of $\vhz$ can only occur in the closed-loop situation. 

Such a decomposition into a linear and memoryless nonlinear parts is called a Lur'e decomposition. Its use is foundational to the study of stability in modern robust control theory and is well explained in two beautiful papers by Zames \cite{Zames:1966a, Zames:1966b}.
The approach to proving global stability in those papers rests on finding conditions where only one pair $(\vf,\vtz)$ is consistent with both open-loop elements. In general, fluid systems are not globally stable and so many solutions may arise. The use of this decomposition first appeared in the analysis of Navier-Stokes systems in \cite{Sharma.ea:2005} and \cite{Sharma:2009} which used sector-bounding arguments applied to stability/control and model reduction respectively.

\section{Derivation of the resolvent operator for the turbulent NSE}

In this section we review how to form the resolvent operator for the incompressible Navier-Stokes equations. There are minor complications arising from the pressure term. In the case where you are able to express the velocity in terms of a divergence-free basis, these difficulties are avoided. The more general case, including pressure, is presented here.

\subsection{The Fourier transformed Navier-Stokes-Equations}

The non-dimensionalised incompressible Navier-Stokes equations are
\begin{align}
    \label{NS}
    \partial_t \vu(\vx, t) + {\vu(\vx, t)} \cdot \nabla {\vu(\vx, t)} =& - \nabla p(\vx, t) + \frac{1}{Re} \nabla^2 {\vu(\vx, t)}\\
    \nabla \cdot {\vu(\vx, t)} =& 0, \nonumber
\end{align}
where $\vx \in X$ is a point in the physical flow domain, $t$ is time, $Re = \rho U L/ \mu$ is the Reynolds number,  $U$ is a characteristic velocity, $L$ is a length scale and $\nabla$ is the gradient operator on $X$. The density of the fluid is $\rho$ and its viscosity $\mu$. Velocity and pressure are thus non-dimensional, with velocity scaled by $U$ and pressure scaled by $\rho U^2$.

As in the general case, first consider the temporal Fourier transform for the state $\vu$,
\begin{align}
    \vhu(\vx, \omega) &= \int_{-\infty}^{\infty} e^{- i \omega t}\vu(\vx, t) \ dt,
    \label{eq:u FT}
\end{align}
From the previous discussion, we know that $\vhu(\vx, \omega=0)$ gives the temporal mean velocity.
Define the fluctuations about the mean as $\vtu(\vx,t)=\vu(\vx,t)-\mean{\vu}$.
Now, treat the pressure similarly to the velocity,
\begin{equation}
    \hp(\vx, \omega) = \int_{-\infty}^{\infty}e^{-i \omega t}p(\vx, t) \ dt.
\end{equation}
Finally, Fourier transform $\vtu(\vx,t)\cdot \nabla \vtu(\vx,t)$ in the same way,
\begin{equation}
    \vhf(\vx, \omega) = \int_{-\infty}^{\infty} e^{-i \omega t} \vtu(\vx, t) \cdot \nabla \vtu(\vx, t) \ dt,
    \label{eq:f FT}
\end{equation}
noticing that $\vhf(\vx, \omega=0)$ yields the time average $\mean{\vtu\cdot \nabla \vtu}$, which does not to have to be zero, even though the time average of $\vtu$ itself is.

Using these definitions and taking the Fourier transform of the Navier-Stokes equations,
\begin{align}
    \int_{-\infty}^{\infty}  e^{-i\omega t} &
    \Big[
       \partial_t \vtu(\vx, t) \nonumber \\
        & + \vtu(\vx, t) \cdot \nabla \vtu(\vx, t) 
        + \mean{\vu}(\vx) \cdot \nabla \mean{\vu}(\vx)  \nonumber \\
        & + \mean{\vu}(\vx) \cdot \nabla \vtu(\vx, t) 
        + \vtu(\vx, t) \cdot \nabla \mean{\vu}(\vx) \nonumber \\
        & + \nabla \left( \mean{p}(\vx) + \tp(\vx, t) \right) \nonumber \\
        & - \frac{1}{Re} \nabla^2 \left( \mean{\vu}(\vx) + \vtu(\vx, t) \right)
    \Big]\ dt = 0. \nonumber
    \label{eq:freq NSE}
\end{align}
This yields both the fluctuation equation and the RANS equation. The former ($\omega \neq 0$) is
\begin{align}
    i \omega \vhu(\vx, \omega) + \mean{\vu}(\vx) \cdot \nabla \vhu(\vx, \omega) + \vhu(\vx, \omega) \cdot \nabla \mean{\vu}(\vx) \\
    = - \nabla \hp(\vx, \omega) + \frac{1}{Re} \nabla^2 \vhu(\vx, \omega) - \vhf(\vx, \omega) \nonumber
    \label{eq:fluctuation}
\end{align}
and the latter ($\omega=0$) is
\begin{equation}
    \mean{\vu}(\vx) \cdot \nabla \mean{\vu}(\vx) = - \nabla \mean{p}(\vx) + \frac{1}{Re} \nabla^2 \mean{\vu}(\vx) - \mean{\vf}(\vx).
    \label{eq:RANS}
\end{equation}
These equations include the pressure. The next steps are in order to eliminate it.

For later convenience, define
\[
    L = \mean{\vu}(\vx)\cdot\nabla + (\nabla \mean{\vu}(\vx))^T
\]
and the Leray projection \cite{Temam:2001}
\[
    \Pi = \left(I - \nabla(\nabla^2)^{-1}\nabla\cdot\right).
\]
We will use the Leray projection to enforce incompressibility allowing the pressure term to disappear. Care must be taken to specify the boundary conditions when inverting the Laplacian.
Using incompressibility, taking the divergence of both sides of the NSE gets rid of the time derivative and gives the pressure Poisson equation, which relates the pressure and the velocity field instantaneously,
\[
    -\nabla^2 p(\vx, t) = \nabla \cdot \left(\vu(\vx, t)\cdot\nabla\vu(\vx, t)\right).
\]
Taking the Fourier transform, and using the Fourier transforms defined earlier, the mean component is
\[
    - \nabla^2 \mean{p}(\vx) = \nabla \cdot (\mean{\vu}(\vx)\cdot\nabla\mean{\vu}(\vx)) + \nabla\cdot \mean{\vf}(\vx)
\]
and at other frequencies,
\begin{equation}
    \nabla^2 \hp(\vx, \omega) = -\nabla \cdot L \vhu(\vx, \omega) - \nabla \cdot \vhf(\vx, \omega).
\end{equation}
Thus, $\hp$ is related to $\vhu$ and $\vhf$ via a linear operator. Substituting these into \eqref{eq:fluctuation}, we get
\begin{equation}
    \left(i\omega I + \Pi L - \frac{1}{Re} \nabla^2\right)\vhu(\vx, \omega)
    = -\Pi \vhf(\vx, \omega).
\end{equation}

\begin{figure}
    \begin{tikzpicture}[auto, looseness=1.2, thick, node distance=2cm,>=latex', decoration={penciline,amplitude=1pt}]

        \node [block] (N) {$\vu\cdot\nabla \vu$};
    \node [block, below of=N] (L0) {RANS};
    \node [block, left of=L0, xshift=-3cm] (FT) {$\int_{-\infty}^{\infty} \vf e^{-i\omega t} \ dt$};
    \node [block, right of=L0, xshift=3cm] (IFT) {$\frac{1}{2\pi} \int_{-\infty}^{\infty} \vhu e^{i\omega t} \ d\omega$};

        \draw [->] (IFT) |- node[near start, right] {$\vu$} (N);
    \draw [->] (N) -| node[near end, left] {$\vf$} (FT);

        \draw [->] (FT) to [in=180, out=0] node[above] {$\mean{\vf}$} (L0);
    \draw [->] (L0) to [in=180, out=0] node[above] {$\mean{\vu}$} (IFT);

    \foreach \n in {2,...,1} {
        \pgfmathsetmacro\q{\n*1.5}
        \node [block, below of=L0, yshift=1cm -\q cm, xshift=\q cm - 3.5 cm] (L\n) {$R(\omega_\n)$};
        \draw [->, dashed] (L0) to [in=90, out=270] node[above, yshift=-1.0cm] {} (L\n);
        \draw [->] (FT) |- node[near end, above] {$\vhf(\omega_\n)$} (L\n);
        \draw [->] (L\n) -| node[near start, above] {$\vhu(\omega_\n)$} (IFT);
    }

    \node [block, below of=L0, yshift=-5cm, xshift=2.5cm] (Ln) {$R(\omega_n)$};
    \draw [->] (L0) to [in=90, out=270] node[above, yshift=-1.0cm] {} (Ln);
    \draw [->] (FT) |- node[near end, above] {$\vhf(\omega_n)$} (Ln);
    \draw [->] (Ln) -| node[near start, above] {$\vhu(\omega_n)$} (IFT);

    \draw [loosely dotted, thick, shorten <= 0.5cm, shorten >= 0.5cm, below of=L0, yshift=-3.5cm, xshift=1cm] (L2) -- (Ln);

\end{tikzpicture}

    \caption{A schematic block diagram, showing the network of resolvent operators and the nonlinear terms which compose the frequency-domain representation of the NSE.}
    \label{fig:NSE resolvent network}
\end{figure}
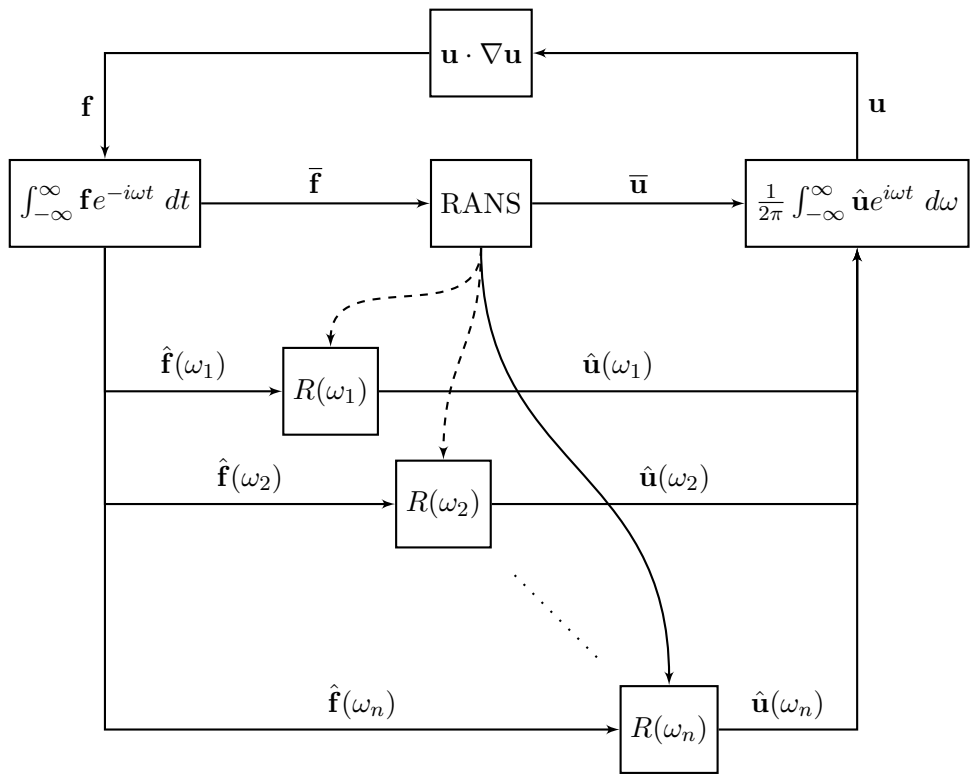

The operator
\begin{equation}
    R(\omega):=-\left(i\omega I + \Pi L - \frac{1}{Re} \nabla^2\right)^{-1}\Pi
\end{equation}
is the resolvent of the NSE linearised about the mean, and
\begin{equation}
    \vhu(\vx, \omega) = R(\omega) \vhf(\vx, \omega).
    \label{eq:resolvent}
\end{equation}
Notice that $R$ depends on the time-average velocity field $\mean{\vu}$ and on frequency $\omega$, and that $\mean{\vu}$ appears naturally and with a consistent interpretation.
The importance of the spectrum of the linear operator about $\mean{\vu}$ thus has a clear interpretation even in a fully nonlinear flow.
The RANS equation \eqref{eq:RANS} and the resolvent equations \eqref{eq:resolvent} are connected via the Fourier transform, as shown in Figure \ref{fig:NSE resolvent network}.
Notice also that $\vf = \vtu \cdot \nabla \vtu$ acts as a memoryless nonlinearity in the sense of Section \ref{sec:Lur'e}, so the earlier discussion on solutions applies.

\section{The application of the singular value decomposition to the resolvent}

We should like to find a sensible basis in which to expand $\vhu$. Ideally, the functions should be orthonormal (to make projections and expansions simple), and chosen and ordered in such a way that a truncation of the expansion should still approximate the true $\vhu$ in a quantifiable way.

Like other modal decompositions including proper orthogonal decomposition \cite{Holmes.Lumley.Berkooz:1996} or dynamic mode decomposition \cite{Schmid:2010}, we will use the singular value decomposition, but on a dynamical flow operator instead of a dataset.

In the case that the dynamics are translation-invariant (such as the streamwise direction in an infinite pipe), the Fourier transform already provides a unitary basis, and we should immediately take the Fourier transform in those directions. In non-invariant directions, such as a wall-normal direction, this is not the case, and we proceed as follows.

Notice that $R(\omega)$ is a linear mapping from the Fourier-transformed `forcing' field to the Fourier-transformed `velocity' field.
Because there is no reason to expect the nonlinear forcing $\vf$ to look like the velocity field, we should expect to use two different bases for $\vhu$ and $\vhf$.
One sensible way to form a basis is the Schmidt decomposition of $R(\omega)$, which provides different bases for $\vhu$ and $\vhf$ and is optimal in useful ways.
With this choice,
\begin{align}
    R(\omega) \vhf(\vx, \omega) = \sum_{j\in \N} \sigma_j(\omega) \inprod{\vhf(\vx, \omega)}{\for_j(\vx, \omega)}_X \res_j(\vx, \omega) ,\\
    \inprod{\res_j(\vx, \omega)}{\res_{j'}(\vx, \omega)}_X = \delta_{j, j'},\nonumber \\
    \inprod{\for_j(\vx, \omega)}{\for_{j'}(\vx, \omega)}_X = \delta_{j, j'}\nonumber,\\
    \sigma_j(\omega) \geq \sigma_{j+1}(\omega).
    \label{svd}
\end{align}
The pairs $\phi_j$ and $\psi_j$ at each $\omega$ are the Schmidt pairs (singular vectors) in the decomposition. The sets of $\phi_j$ and $\psi_j$ at each $\omega$ each form an orthonormal basis (under the inner product on the spatial domain $X$), with basis functions ordered by the singular values $\sigma_j$. This ordering provides a criterion for truncation. Note that the basis is different for each frequency, which is to be expected, since different motions will `resonate' in the flow at different frequencies.

From Section \ref{sec:maximum gain} we know that a particular choice of inner product is implicit in the use of the SVD.
For the incompressible Navier-Stokes equations, calculations involving energy budgets and the nonlinearity are made simpler by the choice of the unweighted spatial $\mathcal{L}_2$ norm, but other systems (such as compressible or reacting flows) require further thought.
Care should be taken when implementing the Schmidt decomposition in numerical codes (i.e.~translating the Schmidt decomposition to a discrete, matrix SVD) to get the weighting matrix associated with the discretisation scheme employed to correspond correctly to the desired inner product.

In fluids applications, $R(\omega)$ often has very large separation between the leading (one or two) singular values and the next.
The physical basis for this is well documented in the literature \cite{McKeon.Sharma:2010} and (in turbulent shear flows) is largely due to a resonance associated with the critical layer. It also turns out that in shear flows, $R$ is non-normal, so in general $\res_j(\vx, \omega) \neq \for_j(\vx, \omega)$.

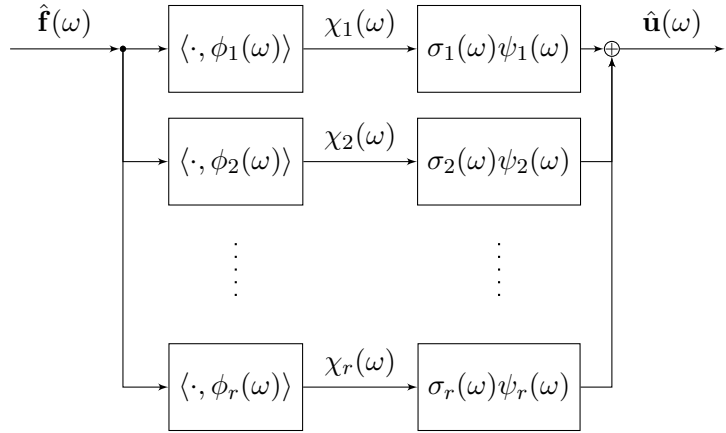
\begin{figure}
    \centering
    \begin{tikzpicture}[auto, node distance=1.5cm,>=latex']
    \node [input] (origin) {};
    \node [node, right of=origin] (demux) {};
    \draw [->] (origin) -- node[above] {$\vhf(\omega)$} (demux);
    \node [sum, right of=demux, xshift=5cm] (sum) {};
    \node [output, right of=sum] (out) {};
    \draw [->] (sum) -- node[above] {$\vhu(\omega)$} (out);

        \node [block, right of=demux] (IPn) {$\inprod{\cdot}{\phi_{1}(\omega)}$};
    \draw [->] (demux) |- (IPn);
    \node [block, right of=IPn, xshift=2cm] (snpsin) {$\sigma_{1}(\omega) \psi_{1}(\omega)$};
    \draw [->] (IPn) -- node[above] {$\chi_{1}(\omega)$} (snpsin);
    \draw [->] (snpsin) -- (sum.west);

        \node [block] (IP2) at (3,-1.5) {$\inprod{\cdot}{\phi_{2}(\omega)}$};
    \draw [->] (demux) |- (IP2);
    \node [block, right of=IP2, xshift=2cm] (snpsin2) {$\sigma_{2}(\omega) \psi_{2}(\omega)$};
    \draw [->] (IP2) -- node[above] {$\chi_{2}(\omega)$} (snpsin2);
    \draw [->] (snpsin2) -| (sum.south);

        \node [block] (IP4) at (3,-4.5) {$\inprod{\cdot}{\phi_{r}(\omega)}$};
    \draw [->] (demux) |- (IP4);
    \node [block, right of=IP4, xshift=2cm] (snpsin4) {$\sigma_{r}(\omega) \psi_{r}(\omega)$};
    \draw [->] (IP4) -- node[above] {$\chi_{r}(\omega)$} (snpsin4);
    \draw [->] (snpsin4) -| (sum.south);

        \draw [loosely dotted, thick, shorten <= 0.5cm, shorten >= 0.5cm] (IP2) -- (IP4);
    \draw [loosely dotted, thick, shorten <= 0.5cm, shorten >= 0.5cm] (snpsin2) -- (snpsin4);

\end{tikzpicture}

    \caption{The NSE resolvent operator, decomposed into into its forcing and response modes and truncated to order $r$. The forcing $\vhf(\omega)$ at frequency $\omega$ is projected onto the forcing modes, to give their scalar coefficients $\chi_i$. These are then amplified by $\sigma_i(\omega)$ and multiplied by the response modes $\psi_i(\omega)$. The superposition of these responses gives the velocity field Fourier coefficient $\vhu(\omega)$.}
    \label{fig:NSE resolvent components}
\end{figure}

The decomposition leads naturally to ordered expansions for both $\vhu$ and $\vhf$ at any particular frequency; expressing each as a weighted superposition of its basis functions gives
\begin{align}
    \vhu(\vx, \omega) &= \sum_{j=1}^\infty \chi_j(\omega) \sigma_j(\omega) \res_j (\vx, \omega), \\
    \vhf(\vx, \omega) &= \sum_{j=1}^\infty \chi_j(\omega) \for_j (\vx, \omega).
\end{align}
The set of $\res_j$ are sometimes called the response modes (or resolvent modes) and the set of $\for_j$ the forcing modes.
At this point, we do not know the scalar coefficients $\chi_i$.
This decomposition truncated up to rank $r$ is illustrated in Figure \ref{fig:NSE resolvent components}. 

If it so happens that $\sigma_1 \gg \sigma_2$, without knowing much about $\vhf$, we may reasonably approximate the Fourier coefficient of $\vhu$ (up to a complex coefficient) by $\vhu(\vx, \omega) \cong \res_1(\vx, \omega)$, regardless of our knowledge of $\vhf$.
By `up to a complex coefficient' it is meant that, while the functional form of each response mode is known, the complex coefficient of the response mode (determining both the phase and magnitude of the wave) is not determined by the decomposition.
However, because the \emph{relative} phase between forcing and response mode pairs is fixed by the decomposition, the phase between different response modes may be fixed either via a direct calculation of the nonlinear forcing, by a projection onto DMD modes, by fitting to a limited set of measurements \cite{Gomez.Blackburn.Rudman.ea:2016b, Beneddine.ea:2017}, or by other methods \cite{Moarref.Jovanovic.Tropp.ea:2014, Moarref.Sharma.Tropp.ea:2013}.

Retaining just the first mode per frequency (a rank-1 approximation) is essentially the same calculation as the `optimal response' found by various authors \cite{Hwang.Cossu:2010}.
The approximation argument does not hold in reverse; which is to say, to approximate $\vhf$ it would require $R^{-1}$ to be approximately rank-1, which is not usually the case.

In some systems, $\chi_j(\omega)$ for the leading few modes may be quite small relative to the following coefficients. This effect may outweigh the effect of any separation of the leading singular values. This may happen, for instance, because the gradient operator involved in calculating the nonlinear term $\vhf$ can act to attenuate the larger scales important at lower frequencies.
In such cases, it is reasonable to approximate $R$ by a higher-rank projection induced by \eqref{svd}, with the number of modes retained determining the level of accuracy, as in \cite{Moarref.Jovanovic.Tropp.ea:2014}.
The extent to which either scenario applies will depend on the particularities of the system under study.

\bibliographystyle{ieeetr}
\bibliography{bibliography}

\begin{thebibliography}{10}

\bibitem{Trefethen.ea:1993}
L.~N. Trefethen, A.~E. Trefethen, S.~C. Reddy, and T.~A. Driscoll,
  ``Hydrodynamics stability without eigenvalues,'' {\em Science}, vol.~261,
  no.~5121, pp.~578--584, 1993.

\bibitem{Jovanovic.Bamieh:2005}
M.~R. Jovanovi{\'c} and B.~Bamieh, ``Componentwise energy amplification in
  channel flows,'' {\em Journal of Fluid Mechanics}, vol.~534, 2005.

\bibitem{Strang}
G.~Strang, {\em Linear Algebra and its Applications}.
\newblock Academic {P}ress, 1976.

\bibitem{Young:1988}
N.~Young, {\em An introduction to Hilbert space}.
\newblock Cambridge {U}niversity {P}ress., 1988.

\bibitem{Curtain.Zwart:1995}
R.~F. Curtain and H.~J. Zwart, {\em An Introduction to Infinite-Dimensional
  Linear Systems Theory}.
\newblock New York: Springer-Verlag, 1995.

\bibitem{Trefethen.Bau:1997}
L.~Trefethen and D.~Bau, {\em {Numerical Linear Algebra}}.
\newblock Society for Industrial and Applied Mathematics, 1997.

\bibitem{Green.Limebeer:1995}
W.~J. Green and D.~J.~N. Limebeer, {\em Linear Robust Control}.
\newblock New Jersey: Prentice Hall, 1995.

\bibitem{McKeon.Sharma:2010}
B.~J. Mc{K}eon and A.~S. Sharma, ``A critical-layer framework for turbulent
  pipe flow,'' {\em Journal of Fluid Mechanics}, vol.~658, p.~336–382, July
  2010.

\bibitem{Zames:1966a}
G.~Zames, ``On the input-output stability of time-varying nonlinear feedback
  systems - {P}art {II}: Conditions involving circles in the frequency plane
  and sector nonlinearities,'' {\em IEEE Trans. on Automatic Control}, 1966.

\bibitem{Zames:1966b}
G.~Zames, ``On the input-output stability of time-varying nonlinear feedback
  systems - {P}art {I}: Conditions derived using concepts of loop gain,
  conicity and positivity,'' {\em IEEE Trans. on Automatic Control},
  vol.~AC-11, pp.~228--238, April 1966.

\bibitem{Sharma.ea:2005}
A.~S. Sharma, D.~J.~N. Limebeer, B.~J. Mc{K}eon, and J.~F. Morrison,
  ``Stabilising control laws for the incompressible {N}avier-{S}tokes equations
  using sector stability theory,'' in {\em Proceedings of the 3rd AIAA Flow
  Control Conference, San Francisco, California}, American Institute of
  Aeronautics and Astronautics, 2005.

\bibitem{Sharma:2009}
A.~S. Sharma, ``Model reduction of turbulent fluids flows using the supply
  rate,'' {\em International Journal of Bifurcation and Chaos}, vol.~19,
  p.~1267–1278, 2009.

\bibitem{Temam:2001}
R.~Temam, {\em {N}avier-{S}tokes Equations: Theory and Numerical Analysis}.
\newblock AMS Chelsea Publishing, 2001.

\bibitem{Holmes.Lumley.Berkooz:1996}
P.~Holmes, J.~L. Lumley, and G.~Berkooz, {\em Turbulence, Coherent Structures,
  Dynamical Systems and Symmetry}.
\newblock Cambridge, U.K.: Cambridge {U}niversity {P}ress., first~ed., 1996.

\bibitem{Schmid:2010}
P.~J. Schmid, ``{Dynamic mode decomposition of numerical and experimental
  data},'' {\em Journal of Fluid Mechanics}, vol.~656, pp.~5 -- 28, 2010.

\bibitem{Gomez.Blackburn.Rudman.ea:2016b}
F.~G\'omez, H.~M. Blackburn, M.~Rudman, A.~S. Sharma, and B.~J. Mc{K}eon, ``A
  reduced-order model of three-dimensional unsteady flow in a cavity based on
  the resolvent operator,'' {\em Journal of Fluid Mechanics}, vol.~798, June
  2016.

\bibitem{Beneddine.ea:2017}
S.~Beneddine, R.~Yegavian, D.~Sipp, and B.~Leclaire, ``{Unsteady flow dynamics
  reconstruction from mean flow and point sensors: an experimental study},''
  {\em Journal of Fluid Mechanics}, vol.~824, pp.~174--201, July 2017.

\bibitem{Moarref.Jovanovic.Tropp.ea:2014}
R.~Moarref, M.~R. Jovanovi{\'c}, J.~A. Tropp, A.~S. Sharma, and B.~J. Mc{K}eon,
  ``A low-order decomposition of turbulent channel flow via resolvent analysis
  and convex optimization,'' {\em Phys. Fluids}, vol.~26, p.~051701, May 2014.

\bibitem{Moarref.Sharma.Tropp.ea:2013}
R.~Moarref, A.~S. Sharma, J.~A. Tropp, and B.~J. Mc{K}eon, ``Model-based
  scaling of the streamwise energy density in high-{R}eynolds-number turbulent
  channels,'' {\em Journal of Fluid Mechanics}, vol.~734, p.~275–316, Oct.
  2013.

\bibitem{Hwang.Cossu:2010}
Y.~Hwang and C.~Cossu, ``Amplification of coherent streaks in the turbulent
  {C}ouette flow: an input--output analysis at low {R}eynolds number,'' {\em
  Journal of Fluid Mechanics}, vol.~643, pp.~333--348, 2010.

\end{thebibliography}

\newpage

\title{Worksheet for resolvent methods}
\maketitle

You have been given a short Matlab function to calculate the Orr-Sommerfeld-Squire operator for a channel.
To keep things simple the operator is formed around the laminar profile.
In the formulation used in the code, the velocity field is Fourier transformed in space, so has streamwise wavenumber $k_x$ and spanwise wavenumber $k_z$,
\[
    \vu(x, y, z, t) \propto \int_{k_x} \int_{k_z} \hat{\hat{\vu}}(y, t; k_x,k_z) \ dk_x dk_z.
\]
As such, a Fourier transform in time will give travelling waves with downstream streamwise wavespeed $-\omega/k_x$.

\begin{exercise}
    Examine the function \texttt{oss.m} and try to understand what it does.
    Find the function return values $A$, $Q$, $C$ and $y$ using resolution $N=150$, streamwise wavenumber $k_x=1$, spanwise wavenumber $k_z=1$, and Reynolds number $Re=1000$.
    The matrix $A$ is the discretised Orr-Sommerfeld-Squire operator, $Q$ is the inner product matrix, $C$ allows calculation of the velocity Fourier coefficients at the wall-normal gridpoints from the state $x$ and $y$ is the gridpoints.
\end{exercise}

\begin{exercise}
    Plot the eigenvalue spectrum of $A$. Look for the eigenvalues closest to the imaginary axis.
    \label{ex:spectrum}
\end{exercise}

\begin{exercise}
    Write a function to find the resolvent of $A$ for a given frequency $\omega$.
    Remember to use the inner product on $Q$ (i.e.~the amplitude being given by $x^*Qx$) for both forcing and response.
    
    Hints:
    \begin{itemize}
        \item You will need the Cholesky decomposition of $Q$, $Q=W^*W$ (Matlab function \texttt{chol}).
        \item You may find it convenient to define a variable $z = Wx$ such that the energy is calculated simply as $z^*z$.
    \end{itemize}
\end{exercise}

\begin{exercise}
    Plot the singular values of the resolvent you calculated. Look at how they decay.
\end{exercise}

\begin{exercise}
    Plot the leading singular value as it changes with $\omega$. Compare the values of $\omega$ where there is the highest gain to the location of the eigenvalues.
\end{exercise}

\begin{exercise}
    Find the leading response mode at $\omega=-1$. Plot the wall-normal velocity's Fourier coefficient as a function of $y$.
\end{exercise}

\begin{exercise}
    These exercises are more time consuming. Try them later.
    \begin{enumerate}
        \item Find the leading resolvent modes without explicitly inverting $i \omega - A$
        \item Explore the relationship between singular value, wavenumber, frequency and the location of the mode peak
        \item Find the same modes using the \texttt{eigs} function
    \end{enumerate}
\end{exercise}

\end{document}